\documentclass[preprint,showpacs,preprintnumbers,amsmath,amssymb]{revtex4}

\pdfoutput=1
\usepackage{graphicx}% Include figure files
\usepackage{dcolumn}% Align table columns on decimal point
\usepackage{bm}% bold math

\newcommand{\be}{\begin{equation}}
\newcommand{\ee}{\end{equation}}
\newcommand{\bear}{\begin{eqnarray}}
\newcommand{\eear}{\end{eqnarray}}
\newcommand{\ba}{\begin{array}}
\newcommand{\ea}{\end{array}}

\newskip\humongous \humongous=0pt plus 1000pt minus 1000pt

\newif\ifdtup

\def\oldreffmt#1{\rlap{[#1]} \hbox to 2\parindent{}}

\def\figfmt#1{\rlap{Figure {#1}} \hbox to 1in{}}

\def\beq{\begin{equation}}
\def\eeq{\end{equation}}
\def\bea{\begin{eqnarray}}
\def\eea{\end{eqnarray}}

\def\bq{\begin{quote}}
\def\eq{\end{quote}}

\relax

\newdimen\tdim
\tdim=\unitlength
\def\bar{\overline}

\begin{document}

\begin{flushright}
ANL-HEP-PR-07-40\\ 
EFI/07-18\\
\end{flushright}

\title{Kaluza-Klein Gluons as a Diagnostic of Warped Models }

\author{ {Ben Lillie$^{b, c}$, \
Jing Shu$^{a, b, c}$, \
Tim M.P. Tait$^{c}$}\\[0.5cm]
\normalsize{$^{a}$ Kavli Institute for Cosmological Physics, University of Chicago, Chicago, IL 60637}\\
\normalsize{$^{b}$ Enrico Fermi Institute and Department of Physics, University of Chicago, Chicago, IL 60637}\\ 
\normalsize{$^{c}$ HEP Division, Argonne National Laboratory, Argonne, IL 60439} }

\begin{abstract}
We study the properties of $g^{1}$, the first excited state of the gluon in 
representative variants of the Randall Sundrum model with the
Standard Model fields in the bulk. We find that measurements of the 
coupling to light quarks (from the inclusive cross-section 
for $pp\to g^{1} \to t\bar t$), the coupling to bottom quarks (from the
rate of $pp\to g^{1} b$), as well as the overall width, can provide 
powerful discriminants between the models.  
In models with large brane kinetic
terms, the $g^1$ resonance can even potentially
be discovered decaying into dijets
against the large QCD background.
We also derive bounds based on existing Tevatron searches for resonant 
$t \bar{t}$
production and find that they require $M_{g^{1}} \gtrsim 950$ GeV.
In addition we explore the pattern of interference between the $g^1$ signal and
the non-resonant SM background, defining an asymmetry parameter for the 
invariant mass distribution. The interference probes the relative 
signs of the couplings of the $g^{1}$ to light quark pairs and to 
$t\bar t$, and thus provides an indication that the top is localized on
the other side of the extra dimension from the light quarks, as is typical
in the RS framework. 
\end{abstract}

%\pacs{11.10.-z, 11.10.Kk, 11.15.-q, 11.25.Mj, 11.25.Uv, 11.30.Rd, 11.40.-q}
\maketitle

\section{Introduction}
\label{sec:intro}

The large hierarchy between the Planck scale where quantum gravity effects
are important, and the scale where the electroweak symmetry is broken, 
drives the wealth of models
at the electroweak scale, and motivates the Large Hadron Collider (LHC)
experiments.  While weakly coupled supersymmetry remains a leading candidate
to stabilize the hierarchy, the Randall-Sundrum (RS) models of a warped extra
dimension \cite{Randall:1999vf} have recently emerged as a fascinating
alternative, which may be connected to string landscape solutions of
the cosmological constant problem \cite{jmr}, and possess an interesting
four dimensional dual interpretation in terms of the composite states
of a strongly coupled nearly conformal field 
theory (CFT) \cite{Arkani-Hamed:2000ds}.

The original RS model had all of the Standard Model confined to the IR
brane (appearing as composites in the dual description).  However, the RS
solution to the hierarchy problem requires only the Higgs to be localized
at the IR boundary, and there are compelling reasons to consider most of the 
SM might actually lie near the UV brane (and thus mostly fundamental with
respect to the CFT in the dual description).  Theories with the SM in the bulk
can incorporate Grand Unification of couplings \cite{Randall:2001gb}, 
motivate the flavor hierarchy of fermion masses 
\cite{Huber:2000ie}, and incorporate a dark matter 
candidate \cite{Agashe:2004ci}.
However, such theories face significant challenge from precision electroweak
observables \cite{Davoudiasl:1999tf}, requiring specific 
features \cite{Agashe:2003zs,Davoudiasl:2002ua,Agashe:2006at}
in order to remain natural.

At the LHC, production of colored states is usually dominant, and the
Kaluza-Klein (KK) excitations of the gluons are particularly attractive,
because they are singly produced and thus have larger rates than the KK
quarks.  Thus, they are usually considered to be likely to be the first
signs of warped physics, and the first
excitation of the gluon ($g^1$) the state for which we will have the most
statistics available in order to unravel the details of the underlying
theory.  They are the natural place to explore whether or not we can
use LHC data to determine which particular detailed RS model has been
realized in nature, and which parameters describe it.
Recently, significant work has begun on some of 
the simplest RS constructions to study the production and decay
of the first KK mode of the gluon, in order to determine the reach
of the LHC to discover RS through its detection \cite{Agashe:2006hk}.

While the KK gluon is the most promising avenue to discover RS, it is
nevertheless challenging.  The coupling to the light quarks that are the
primary constituent of the proton, while characterized by the strong
coupling, are somewhat suppressed by the localization of the light fermions
close to the UV boundary (in the CFT language, the light fermions 
are largely fundamental
fields and couple to the gluon largely through a small mixing with CFT states).
This leads to somewhat smaller production cross sections than are typical
of QCD.
The decay of the gluon is expected to be predominantly into top quarks, 
a consequence of the large top mass, which necessitates that top is itself 
located close to the IR brane (mostly composite).  The tops are produced 
from a very heavy resonance, and are highly boosted, which makes it
experimentally challenging to reconstruct them from the large QCD
backgrounds \cite{Agashe:2006hk}.

In this article, we explore several more of the commonly considered theories
which attempt to render RS consistent with precision electroweak data.
We consider the model with a simple $SU(2)$  \cite{Agashe:2003zs}
custodial symmetry (already
studied before \cite{Agashe:2006hk}) as a beginning, and also consider
models with large brane kinetic terms \cite{Davoudiasl:2002ua} or an expanded
custodial symmetry which protects the $Z$-$b$-$\bar{b}$ vertex from
large corrections \cite{Agashe:2006at,Contino:2006qr,Carena:2007ua} in
order to characterize the difference in the properties of the first KK mode
of the gluon in each case.  We find that there are general features which
can discriminate between the cases, and thus that the specific realization
of the RS model leaves an imprint in the properties of the KK gluon.

The article is laid out as follows: in section~\ref{sec:models}, we review
the specific details of the models under consideration.  In 
Section~\ref{sec:xsec}  we show the $g^1$ production rates
and decay properties and show how the strong coupling
can lead to interesting finite width effects in section~\ref{sec:interfere}.
Section~\ref{sec:conclusions} contains our conclusions.

\section{Models}
\label{sec:models}

\subsection{The Basic RS Model with the SM in the Bulk}

The basic RS model is a slice of AdS$_5$ with the background metric
\begin{equation}
ds ^2 =  \Big( \frac{z_h}{ z}\Big)^2  \left[ \eta_{ \mu \nu } 
d x ^{\mu } d x^{ \nu } + ( d z ) ^2  \right] , \label{metric-conf}
\end{equation}
with curvature $\kappa  = 1/z_h \lesssim M_{Pl}$.
$x^\mu$ are the coordinates of the
four large dimensions, $z$ parameterizes the coordinate
along the extra dimension, and $\eta_{\mu \nu} = Diag(-, +, +, +)$ is
the four-dimensional metric.  Greek letters denote the four large
dimensions $0,1,2,3$ and capital roman letters include the fifth dimension as
well.
The UV boundary is at $z_h = 1/ \kappa$
where the scale factor $(z_h / z)^2 = 1$ and the IR boundary is at
$z_v \sim 1 / {\rm TeV}$, as motivated by the hierarchy problem.

We are particularly interested in a model where all Standard Model
(SM) fields, except perhaps the Higgs, propagate in the entire 5-d spacetime,
and will be primarily concerned with the gluon and colored fermion
fields. The action for the gauge fields and fermions is,
\bea
S = \int d^5x \sqrt{-g} \left\{-\frac{1}{4g_5^2} F_{MN}^{a}F^{MN \, a}
+ i \overline{\Psi} \Gamma^{\dot{M}} e_{\dot{M}}^{M} 
D_M \Psi + i c \kappa \overline{\Psi} \Psi 
\right\} .
\label{eq:action0}
\eea
where $\Gamma^{\dot{M}}$ 
are the 5d ($4 \times 4$) Dirac matrices, $e_{\dot{M}}^{M}$ is the veilbein,
$a$ is an adjoint gauge index and $c$ parameterizes the magnitude
of a bulk mass for the fermion in units of the curvature.

We work in a unitary gauge $A_5 = 0$, and decompose the 5d fields in KK modes,
\bea
A^a_\mu (x,z) = \sum_n A^{a(n)}_\mu(x) g^{(n)}(z) \, , \\
\Psi_{L,R} (x,z) = (\kappa z)^{3/2} \sum_n \psi^{n}_{L,R}(x) 
\xi^{(n)}_{L,R} (z) ~.
\eea
The wave functions are given by combinations of Bessel functions
\bea
g^{(n)}(z) & = & N_n
\: z \: \left[ J_1(m_n z)+ b_n Y_1(m_n z)\right].
\label{eq:KKgluon}
\eea
with normalization factor $N_n$ and
admixture controlled by $b_n$.  The mass spectrum is controlled by the
boundary conditions, with the masses satisfying,
\bea
b_n = - \frac{J_0 \left( m_{n} z_h \right)}{Y_0 \left(  m_{n} z_h  \right) } 
=  - \frac{J_0 \left(  m_{n} z_v \right)}{Y_0 \left( m_{n} z_v \right)} ~.
\label{eq:b0}
\eea
For an unbroken gauge group, there is a zero mode with wave function
$g^0(z) = 1 / \sqrt{L}$, $L = 1/k \log z_v / z_h$.  Of particular note 
for the following is the fact that the light KK states
have most of their support close to the IR boundary.

The physics of bulk fermions was worked out in \cite{Grossman:1999ra}. The spectrum depends sensitively on the bulk mass term $c$.  To remove
unwanted light degrees of freedom, we impose the boundary conditions such
that either the right- or the left-chiral zero mass component is projected
out.  The KK states form left- and right-chiral pairs whose wave functions
are also Bessel functions, 
\bea
\xi^{(n)}_{\pm} (z) =  {\cal N}_n (\kappa z) \left[ 
J_{|c\pm 1/2|}\left( m_n z \right) 
+ \beta_n Y_{|c\pm 1/2|} \left( m_n z\right) \right], \label{eq:fermion}
\eea
where $-~(+)$ are for the right- (left-) chiral modes, and
the masses are determined by imposing the equality,
\bea
\beta_n & = & \frac{J_{|c - 1/2|}\left( m_n z_h \right)}
{Y_{|c - 1/2|}\left( m_n z_h \right)} 
= \frac{J_{|c - 1/2|}\left( m_n z_v \right)}
{Y_{|c - 1/2|}\left( m_n z_v \right)}
~,
\eea
and ${\cal N}_n$ is a normalization factor.
The zero mode wave functions are,
\bea
\xi^{(0)}(z) = {\cal N}_0 \left( \kappa z \right)^{1/2- c_{\Psi} } ~.
\eea
These wavefunctions assume the right-handed zero mode is the one allowed by the boundary conditions; the 
left-handed case is given by $c \rightarrow -c$.  The zero mode is
exponentially peaked  toward the UV boundary for $c < -1/2$ and
toward the IR for $c > -1/2$.  To avoid confusion, we adopt a notation 
where $c$'s explicitly refer to right-chiral fields, so the left-chiral 
fermions should be understood to actually have $-c$ as their mass parameter.

Assuming $\mathcal{O}(1)$ 5D Yukawa couplings, the hierarchy in the
effective 4D Yukawa couplings can be motivated by the exponential suppression
of the wave functions at the IR boundary for order one differences in $c$.
In particular, one cannot allow strong suppression of the top quark wave
functions on the IR boundary, because to reproduce the observed top mass
one would have to adjust the 5d Yukawa coupling to be too strong to
have a perturbative description.  There is further motivation from
precision electroweak data \cite{Agashe:2003zs,Davoudiasl:2002ua}, which
prefers the light fermion mass parameters to be close to $-1/2$ 
(including the left-handed top, as it comes along with the left-handed
bottom, leading to tension in the choice of $c$ for $Q_3$) in order
to cancel the leading contribution to the $S$ parameter from the weak boson
KK modes.  With this setup, and the additional suppression of the $T$ parameter
from a custodial $SU(2)$, masses of the KK modes of around $3$ TeV
are roughly consistent with precision measurements.

Specifically, we consider $c_{t_R} \sim  0$, $c_{Q_{3L}} \sim 0.4$,
and all others $c_f \lesssim -0.5$.  As we will see shortly, the physics 
we study does not depend strongly on $c$ once it is $< -1/2$, so the 
specific values for light fermions
are not important.  The choices of $c$ specify the fermion
zero mode wave functions, and we compute the couplings of the first KK gluon
to the fermion zero modes as the integral over the wave functions.
The light quarks all have very similar couplings of roughly
$g_{f} \simeq -g_S / 5$, the third family left-handed quarks
$g_{Q3} \simeq g_S$, and the right-handed top quark $g_{t} \simeq 4 g_S$,
where $g_S$ is the strong coupling constant which characterizes the coupling
of the gluon zero mode.

\subsection{IR Brane Kinetic Terms}

An alternate way to render precision electroweak data consistent with
low KK mode masses is to include large-ish kinetic terms for the gauge
fields on the IR brane \cite{Davoudiasl:2002ua}.  
Such terms repell the KK mode wave functions from the brane, and have a 
large effect on the phenomenology of the KK modes \cite{Carena:2002me}.  
Brane terms are a class of higher dimensional operators of the 5d theory,
\bea
-\frac{1}{4g_5^2} \int d^5x \sqrt{-g} \left\{ F_{MN}^{a}F^{MN \, a}\right\}
2 \: r_{IR} \: \delta \left( z - z_v \right)
\label{eq:actionbk}
\eea
and will be induced by the orbifold boundary conditions and localized
fields \cite{Georgi:2000ks}.  
Their magnitude $r_{IR}$ is a free parameter of the effective theory.  
While the
size of the IR boundary kinetic term for the gluon is not closely connected
to the quality of the electroweak fit, one would expect that if the UV physics
is such that there are large IR kinetic terms for the electroweak bosons,
such terms are probably also large for the gluon as well.  Thus, if one
could infer the presence of large gluon terms, it would at least suggest that
a similar term is present in the electroweak sector, and responsible for the
success of the SM in explaining the electroweak fit.

\begin{figure}
\begin{tabular}{cc}
\hspace*{-1cm}
\includegraphics[angle=0,scale=0.45]{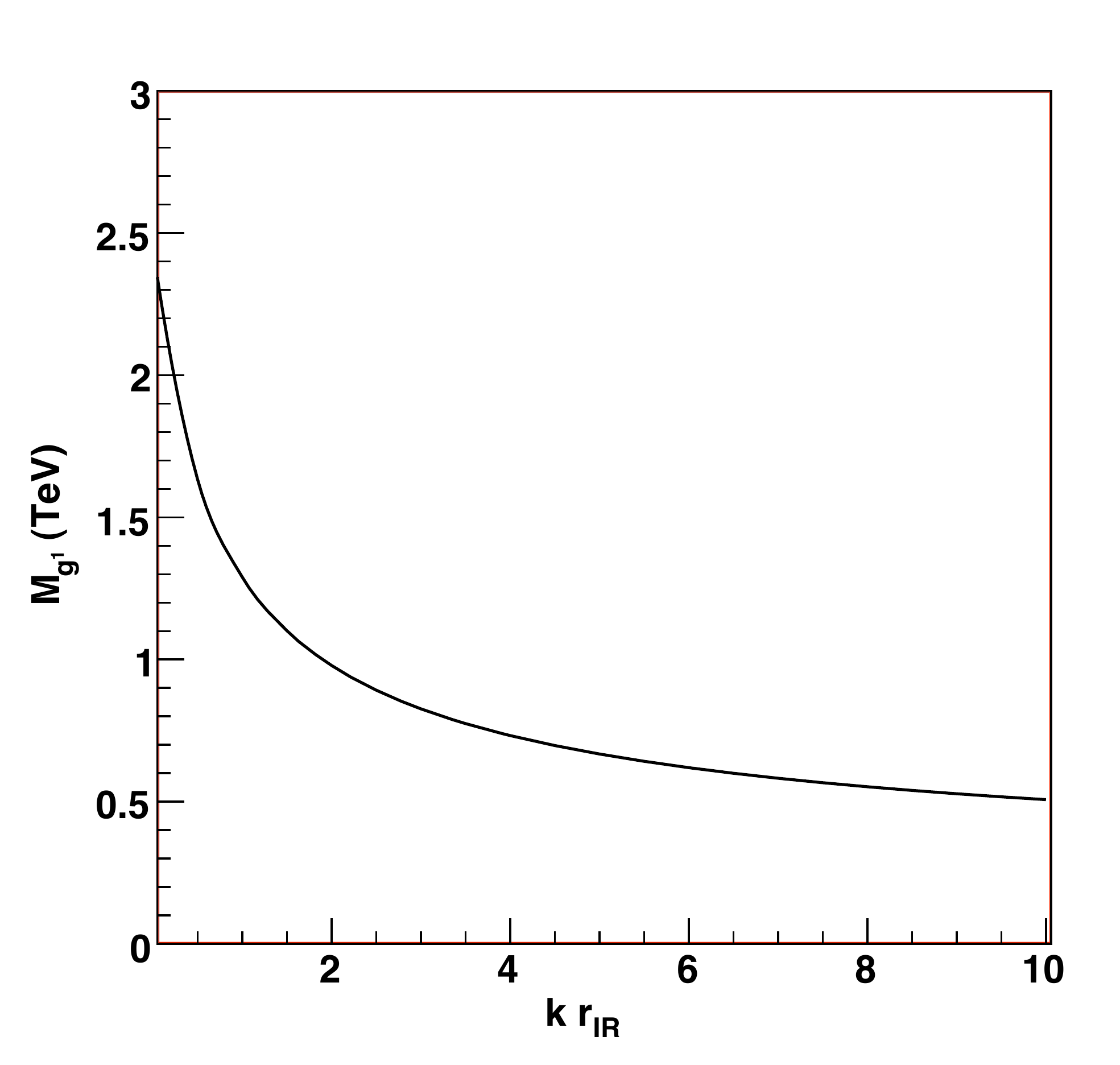} &
\includegraphics[angle=0,scale=0.45]{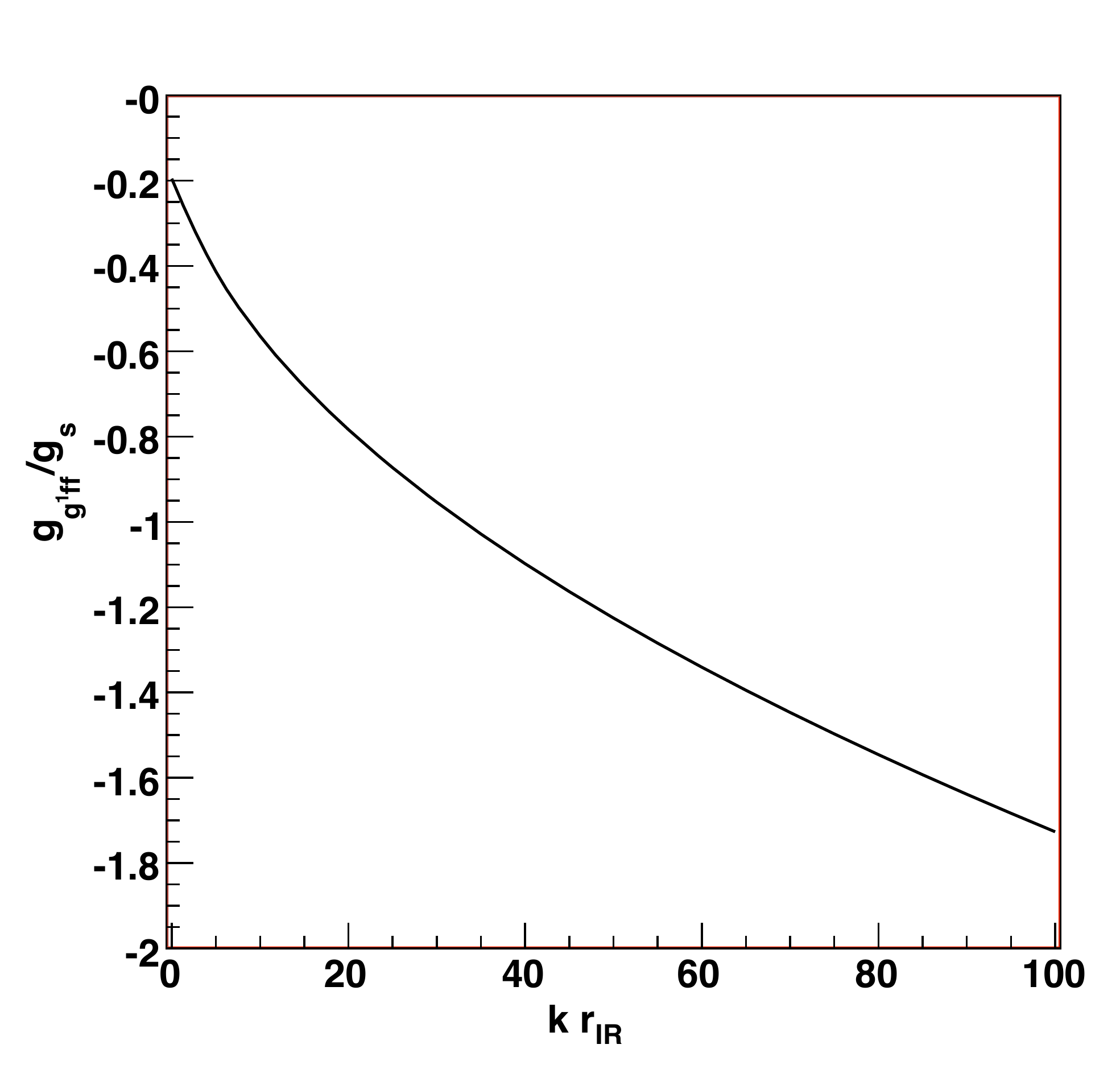} \\
\end{tabular}
\caption{\label{fig:bkm}The 1st KK gluon mass in units of $1/z_v$  and 
coupling of the first KK gluon to a fermion zero mode localized 
at UV brane as a function of brane kinetic term $\kappa r_{IR}$.}
\end{figure}

\begin{figure}
\hspace*{-0.75cm}
\includegraphics[angle=0,scale=0.65]{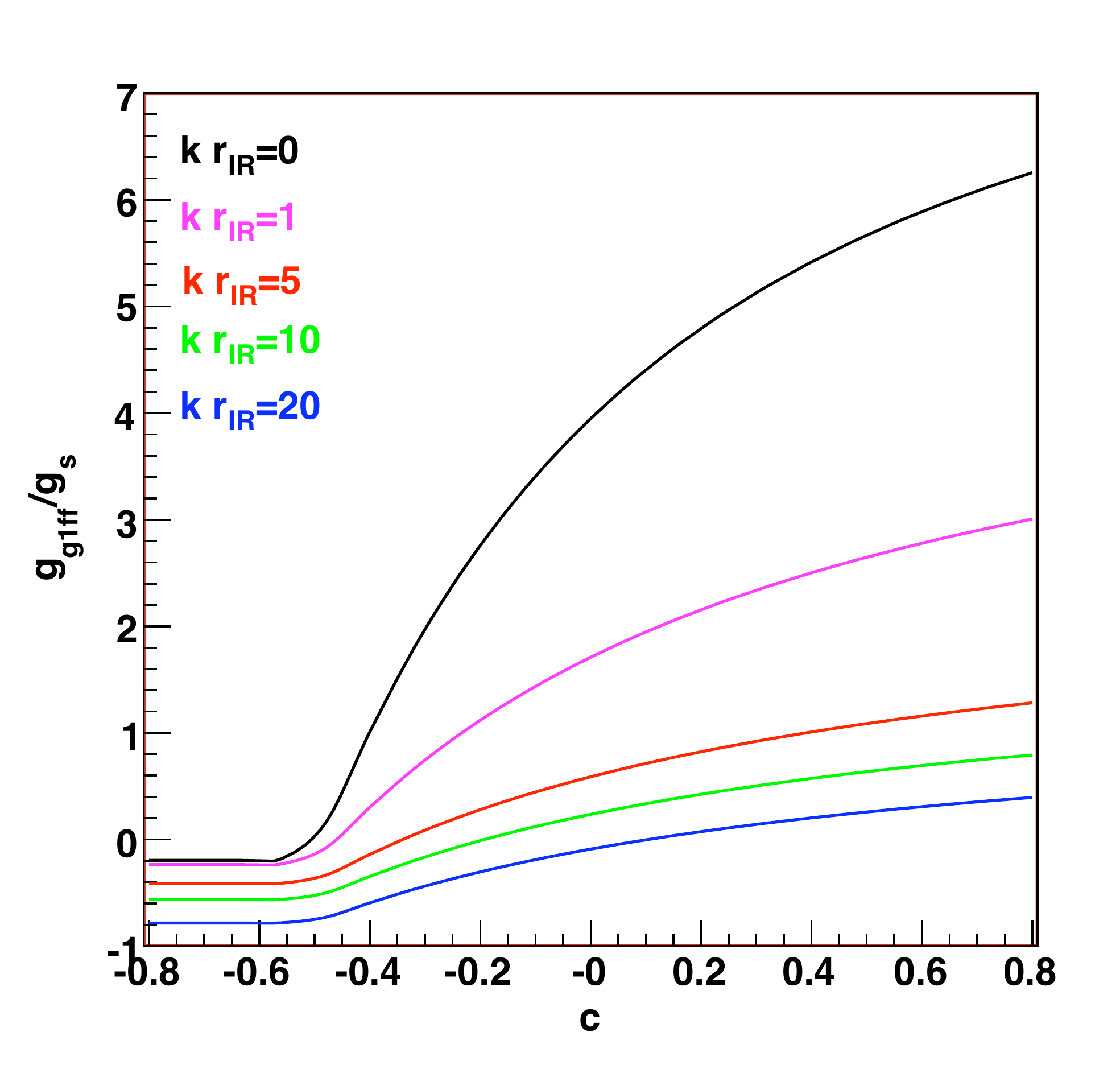} 
\caption{\label{fig:gffbkcoup} 
Coupling of the first KK gluon (with respect to the zero mode gluon coupling)
with $\kappa r_{IR} = 0,1,5,10,20$ (descending) 
to a fermion zero mode as a function of bulk mass parameter $c$. }
\end{figure}

The IR boundary kinetic term does not affect the form of the bulk wave
functions, Eq.~(\ref{eq:KKgluon}).  The boundary conditions become
\bea
b =  - \frac{J_0 \left( m_{n} z_h \right)}{Y_0 \left(  m_{n} z_h  \right) } 
= - \frac{J_0 \left(  m_{n} z_v \right) 
- (\kappa \,r_{IR})  m_{n} \, z_v \: J_1 \left( m_{n} z_v \right)}
{Y_0 \left( m_{n} z_v \right)
- (\kappa \,r_{IR})  m_{n} \, z_v \: Y_1 \left( m_{n} z_v \right) } ,
\label{eq:eigenmassbk}
\eea
indicating different admixture of the Bessel functions $J_1$ and $Y_1$
in the solutions. While there is no analytic solution for the masses, 
they may be easily obtained numerically.  In Figure~\ref{fig:bkm},
we show the variation of the first KK mode gluon mass and coupling
to UV-localized states as a function of the magnitude of the IR brane term
$\kappa r_{IR}$.  In Figure~\ref{fig:gffbkcoup}, we show the dependence of
the coupling on $c$ for a few different choices of $\kappa r_{IR}$.
The inclusion of the boundary terms ameliorates the strongest constraints
from precision electroweak data, and opens up considerably more freedom
to choose the fermion $c$'s.  However, in computing properties below, 
we imagine a situation in which the $c$'s are as in the $SU(2)$ custodial
version outlined above (for example, to explain the flavor hierarchies), 
with large contributions to the electroweak
$T$-parameter controlled by the IR boundary kinetic terms.

\subsection{Holographic Higgs with Expanded Custodial Symmetry}

The models with a custodial $SU(2)$ symmetry or large IR boundary kinetic
terms (combined with the choices of the $c$'s motivated above) continue to
be challenged by the large top mass, which we saw did not allow $Q_3$
to be pushed quite as far away as was optimal for the lighter
fermions.  This results in corrections to the $Z$-$b_L$-$\bar{b}_L$
coupling compared to those of light fermions
which are slightly too large for the experimental errors, and
push in a direction unhelpful for $A^{FB}_b$ \cite{Choudhury:2001hs}.

In \cite{Agashe:2006at}, it was noticed that a subgroup of the custodial 
symmetry can protect the $Z$-$b_L$-$\bar{b}_L$ coupling, provided the third 
generation doublet is embedded in a representation for which the $SU(2)_L$ and
$SU(2)_R$ representations (and the third component of each) are the same.
This implies that to better fit $Z$-$b_L$-$\bar{b}_L$, we expand $Q_3$ into
a bi-doublet under ($SU(2)_L$, $SU(2)_R$).  The unwanted additional fermions
in the bi-doublet are removed from the zero mode spectrum by adjusting their
boundary conditions.  Having promoted $Q_3$ to a bi-doublet,
we recover freedom to consider the $c$ parameter for $Q_3$ 
very different from $-1/2$.  

In order to have a specific framework, 
we analyze the model of gauge-Higgs unification \cite{Carena:2007ua}
(similar to an earlier model \cite{Agashe:2004rs})
in which the allowed parameter space is analyzed in great detail
\cite{Carena:2006bn}, reproducing light fermion masses and mixings,
and demanding consistency with flavor-changing neutral currents induced
by the KK modes of the gauge bosons.  While some of the features are
particular to the gauge-Higgs unified model and the mechanism by which it
realizes fermion masses and mixings, some of the most important features
are fairly generic to models in which an expanded custodial symmetry
is protecting $Z$-$b_L$-$\bar{b}_L$.

The bulk gauge symmetry is $SO(5) \times U(1)_X$, broken by boundary
conditions to $SU(2)_L \times SU(2)_R \times U(1)_X$ on the IR boundary, 
and to the Standard Model $SU(2)_L \times U(1)_{Y}$ gauge
group on the UV brane~\cite{Agashe:2004rs}. The $U(1)_{X}$
charges are adjusted so as to recover the correct hypercharges,
where $Y/2 = T^{3}_{R} + Q_{X}$ with $T^3_{R}$ the third $SU(2)_{R}$
generator and $Q_{X}$ the $U(1)_{X}$ charge.  As motivated above, we wish
$Q_3$ to be part of a bi-doublet, and an economical choice is to embed it
in a ${\bf5}_{2/3}$ of $SO(5)$ (the subscript refers to the $U(1)_X$ charge). 
As discussed in \cite{Carena:2006bn}, it is preferable to place $t_R$
in a seperate ${\bf 5}_{2/3}$ to avoid large negative corrections to the
$T$ parameter.  $b_R$ is part of a ${\bf 10}_{2/3}$, allowing for the
bottom Yukawa coupling, and the first and second generations are
replicas of this structure in order to generate CKM mixing in a
straight-forward way.  Enhanced coupling to bottom quarks is also 
potentially a signal of RS attempts to explain the observed deviation 
in $A_{FB}^b$ \cite{Djouadi:2006rk}.

The scan over parameters of \cite{Carena:2007ua} prefers that
the quarks and leptons of the first two generations are localized close 
to the Planck boundary in order to suppress flavor changing neutral currents.
 The expanded custodial symmetry, combined with relatively light KK modes
for the $Q_3$ custodial partners, is so efficient at suppressing
contributions to the $T$ parameter, that it reduces some of the usual SM
top contribution, and can result in $T$ large and negative, in conflict
with the electroweak fit \cite{lepewwg}.  To ameliorate this new
concern, the freedom to consider $Q_3$ closer to the IR boundary 
(compensated by moving $t_R$ somewhat away from it) is crucial,
allowing $c_{Q_{3L}} \sim 0.2$, $c_{t_R} \sim -0.49$, and 
$c_{f} \lesssim -0.5$, for which the couplings to the first KK mode of the
gluon are approximately $g_f \sim -g_S /5$, $g_{t} \sim 0.07 g_S$,
and $g_{Q3} \sim 2.76 g_S$.

This model generically leads to very light KK quarks, the lightest of which
are the $SO(5)$ bi-doublet partners of the right-handed 
up-type quarks of the first two generations $u_i$ (by virtue of the choice
of $c$ for the two light generations)
\cite{Carena:2007ua}.
Each generation contains
\begin{eqnarray}
\begin{array}{c}
\begin{array}{ccccccc}
~ & ~ & Q^{i}_{2R} &=& \begin{pmatrix}
\chi^{u_{i}}_{2R}(+,-) & q^{\prime {u_{i}}}_R(+,-) \\
\chi^{d_{i}}_{2R}(+,-) & q^{\prime {d_{i}}}_R(+,-) \end{pmatrix},
\end{array}
\end{array}
\label{multiplets}
\end{eqnarray}
along with their $(-,+)$ left-handed Dirac partners.
The $(\pm,\,\mp)$ refers 
to their boundary conditions on the (UV, IR) boundaries, 
and do not lead to zero modes (as desired), and modify the
equation which determines their masses and admixture of Bessel functions.
For the right-handed $(+,-)$ states this leads to,
\begin{eqnarray}
\beta_n & = & 
\frac{ J_{ |c-1/2| } \left( m_n z_v\right) }
{ Y_{|c-1/2| } \left( m_n z_v \right) } 
= \frac{ J_{|c-1/2| \mp 1} \left( m_n z_h \right)  } 
{ Y_{ |c-1/2| \mp 1} \left( m_n z_h \right) } ~,
\label{keyequation}
\end{eqnarray}
with upper(lower) signs for $c>-1/2$~($c<-1/2$).
The left-handed $(-,+)$ states satisfy,
\begin{eqnarray}
\beta_n & = & 
\frac{ J_{ |c+1/2| } \left( m_n z_h \right) }
{ Y_{|c+1/2| } \left( m_n z_h \right) } 
= \frac{ J_{|c+1/2| \mp 1} \left( m_n z_v \right)  } 
{ Y_{ |c+1/2| \mp 1} \left( m_n z_v \right) } ~.
\label{keyequation2}
\end{eqnarray}

\begin{figure}
\begin{tabular}{cc}
\hspace*{-1cm}
\includegraphics[angle=0,scale=0.45]{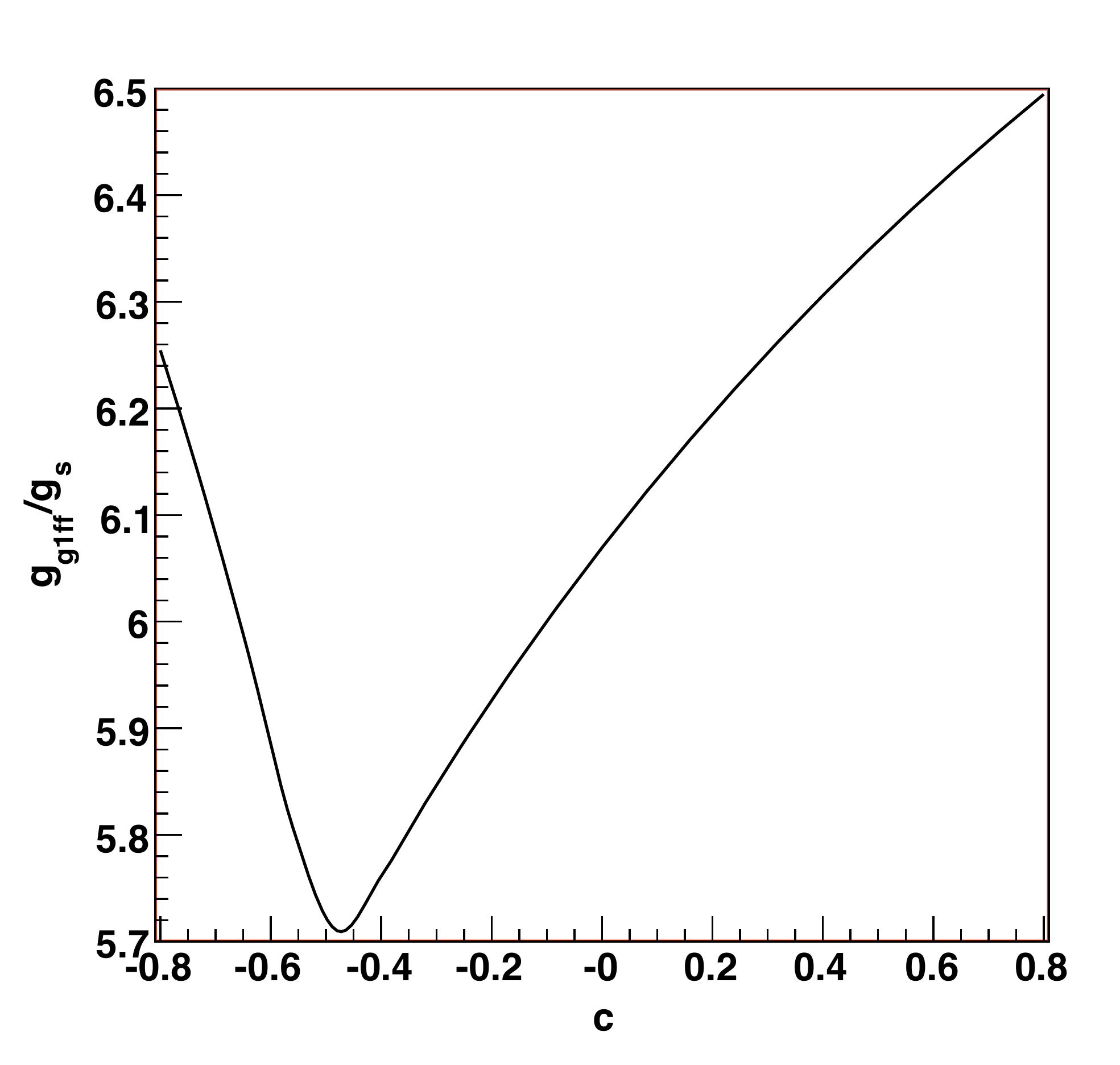} &
\includegraphics[angle=0,scale=0.45]{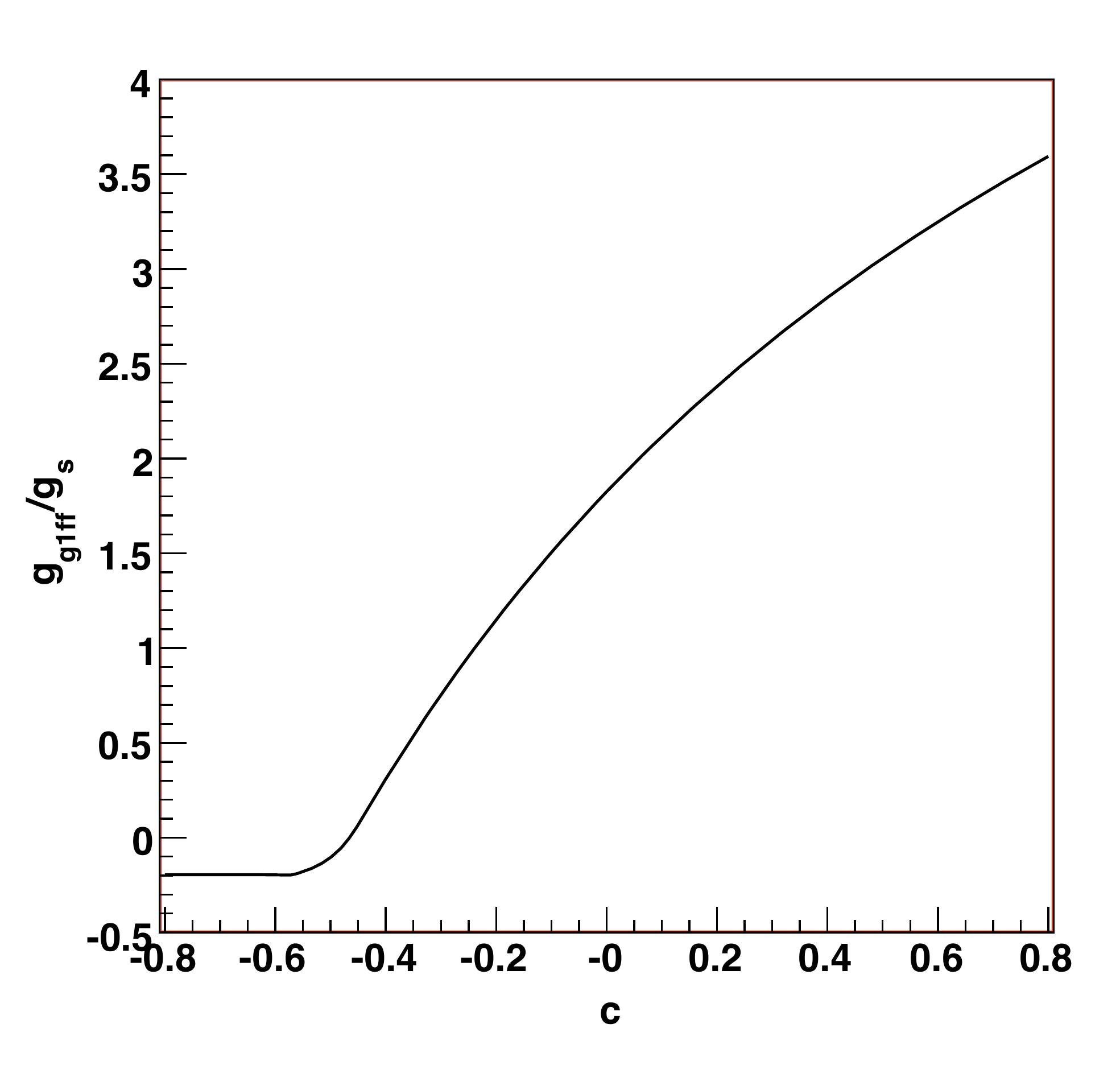} \\
\end{tabular}
\caption{\label{fig:gffmpcoup}
Coupling of the First KK mode of the gluon to the
light KK modes of the
custodial partners of the right-handed up-type quark as a function of the
bulk mass parameter $c$.  The left panel shows the left-handed coupling
whereas the right panel the right-handed coupling.}
\end{figure}

Armed with these wave functions, we compute the coupling of these potentially
light first KK modes of the custodial partners
to the first KK mode of the gluon.  The results for both
chiralities are presented in Figure~\ref{fig:gffmpcoup}, and indicate that
one chirality is always
very strongly coupled, $g \sim 6 g_S$, irrespective of the value of $c$.

\subsection{A Warped Higgsless Model}

A final variant of the warped theory has no Higgs, and breaks the electroweak
symmetry by boundary conditions \cite{Csaki:2003dt}.  The need for the KK modes of the
weak vector bosons to unitarize $WW$ scattering implies that the scale of
KK mode masses is at most several hundred GeV, whereas the need to be
consistent with precision electroweak data and realize a large
top mass requires \cite{Cacciapaglia:2006mz}\footnote{We would like to thank Giacomo Cacciapaglia for assisting us with determining these couplings for the model of \cite{Cacciapaglia:2006mz}}
\bea
 g_{t} = 2.5 g_S ~,\ \  
g_{Q3} = 2 g_S ~,\ \ 
g_{b} = -0.32 g_S ~,\notag \\  
g_{\rm{other RH}} = -0.33 g_S ~, \ \ 
g_{\rm{other LH}} = 0.15 g_S .
\label{eq:hlfermion}
\eea
We see that the basic trend is very similar to the other RS models we consider.
The main distinguishing feature is the fact that the mass of the KK gluon
must be less in order for the Higgsless model to remain consistent with
perturbative unitarity.

\section{Production and Decay}
\label{sec:xsec}

\begin{figure}[tbp]
\hspace*{-0.75cm}
\includegraphics[angle=0,scale=0.65]{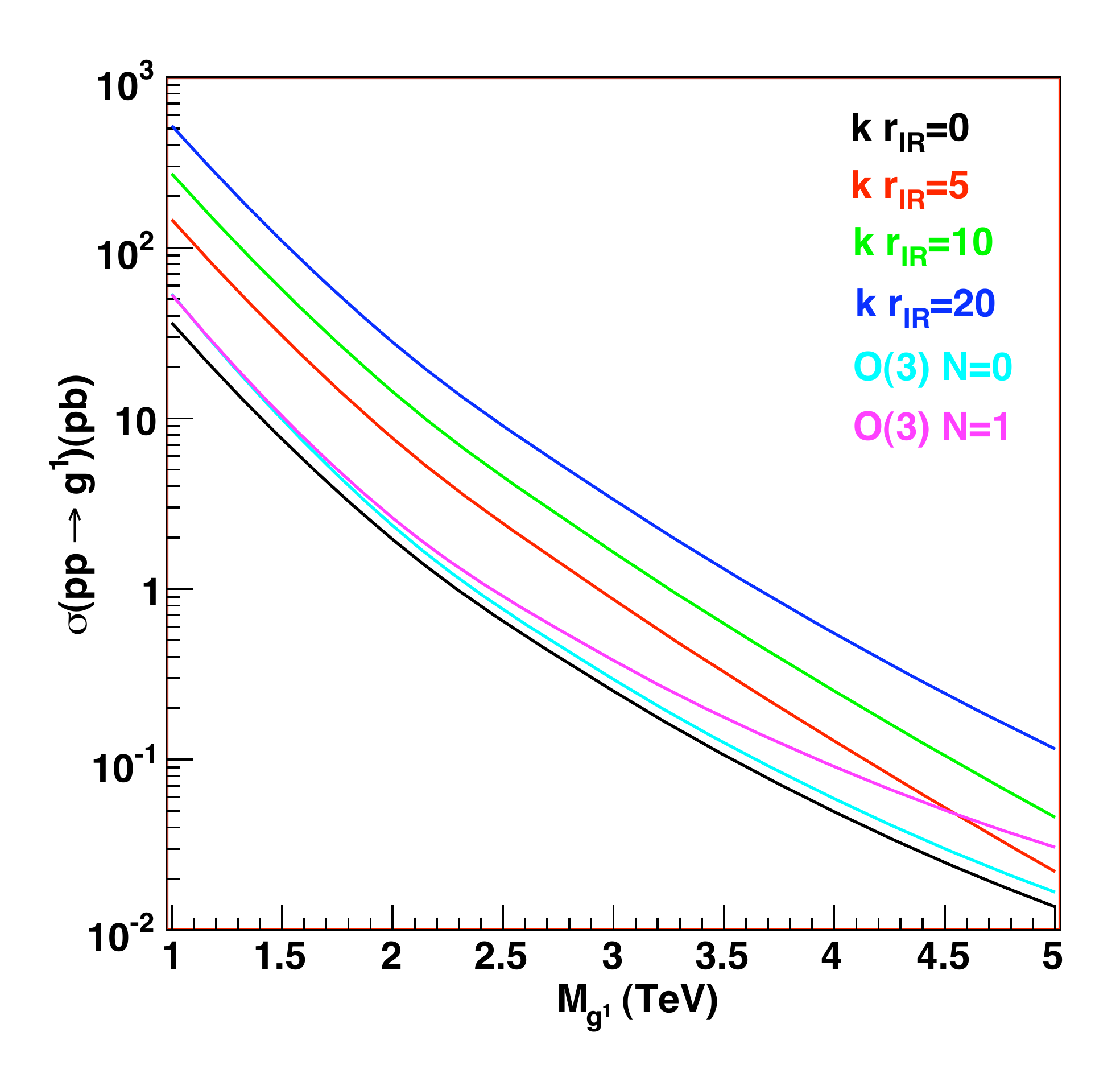} 
\caption{\label{fig:xsec} 
Cross section for $pp \rightarrow g^1$ at the LHC,
for standard RS with the SM in the bulk ($\kappa r_{IR}=0$), three models
with large brane kinetic terms ($\kappa r_{IR} = 5, 10, 20$) and the model
with a larger custodial symmetry, in the cases when $N=0$ or $1$, of the
additional KK custodial partner quarks are light enough that $g^1$ can
decay into them.}
\end{figure}

The details of production of KK gluons at the LHC
will depend on how they couple to the relevant partons at LHC energies, and these differences will give us a powerful way to discriminate between models. Note that the vertex with two gluons and a KK gluon is zero at tree-level, meaning that the dominant production mode is $q\bar q$ annihilation.
As is well-known, in the standard RS framework the KK gluon
coupling to all fermions aside from $t_R$ is suppressed.  As we saw above,
models with brane kinetic terms can increase couplings to UV-localized fields,
which increases the rate and affect the branching ratios.  In addition, the
models with custodial symmetry have a large coupling to $Q_3$, turning on
a new production mode from bottom fusion, but have a smaller branching
ratio because the decay into the custodial partner KK modes may compete
with top.  In Figure~\ref{fig:xsec}, we plot the 
cross section, calculated at leading order by MADGRAPH \cite{Alwall:2007st},  
$pp \rightarrow g^1$ for $\sqrt{S} = 14$~TeV,
as a function
of $g^1$ mass for the models considered above, including the channels
initiated by light quark fusion and bottom fusion.

\begin{figure}[tbp]
\hspace*{-0.75cm}
\includegraphics[angle=0,scale=0.65]{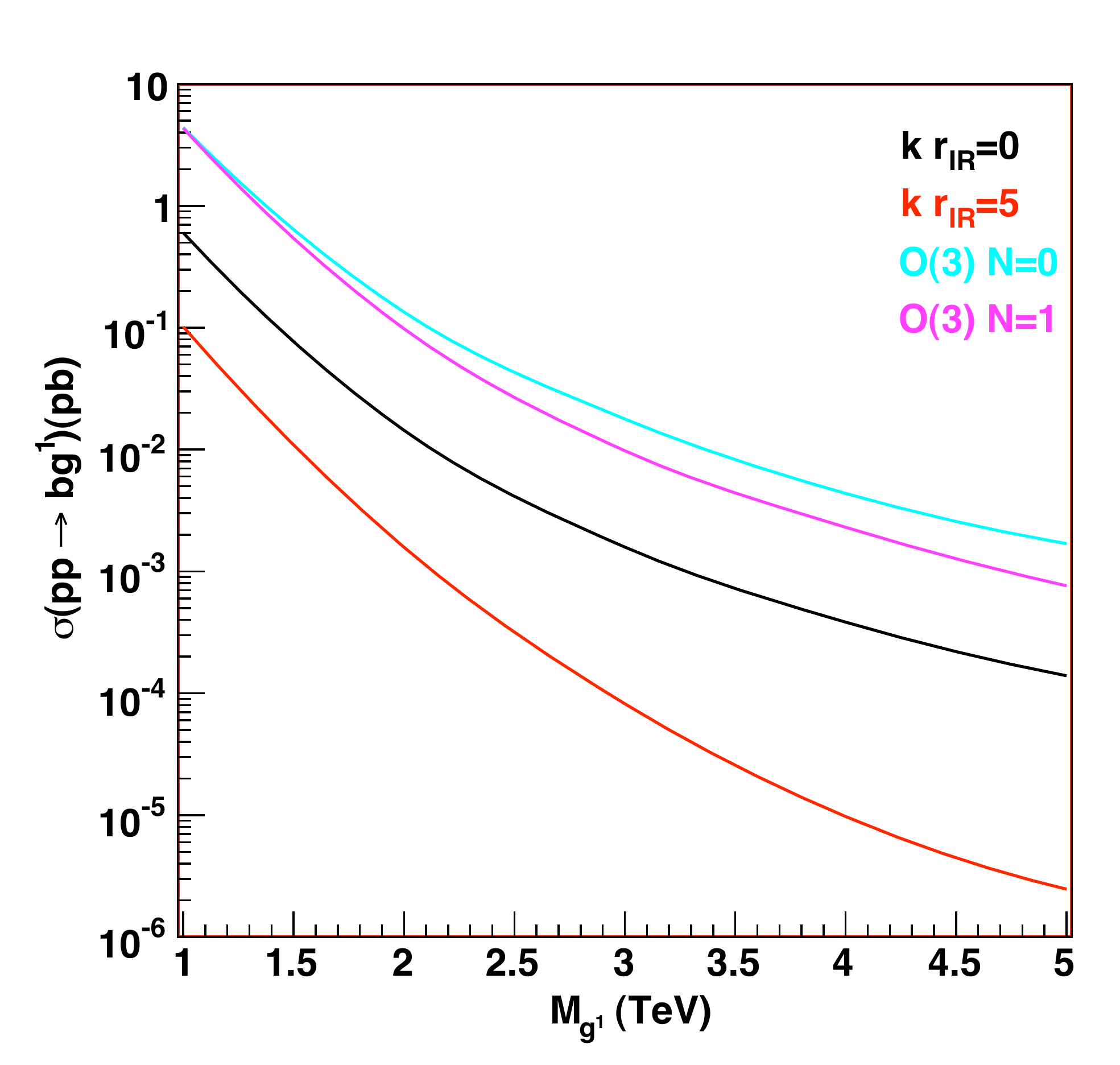} 
\caption{\label{fig:bxsec} 
Cross section for $pp \rightarrow b g^1$ at the LHC,
for standard RS with the SM in the bulk ($\kappa r_{IR}=0$), a model
with a large brane kinetic term ($\kappa r_{IR} = 5$) and the model
with a larger custodial symmetry, in the cases when $0$ ($1$) of the
additional KK custodial partner quarks are light enough that $g^1$ can
decay into them.}
\end{figure} 

\begin{figure}[tbp]
\hspace*{-1.8cm}
\includegraphics[angle=0,scale=0.85]{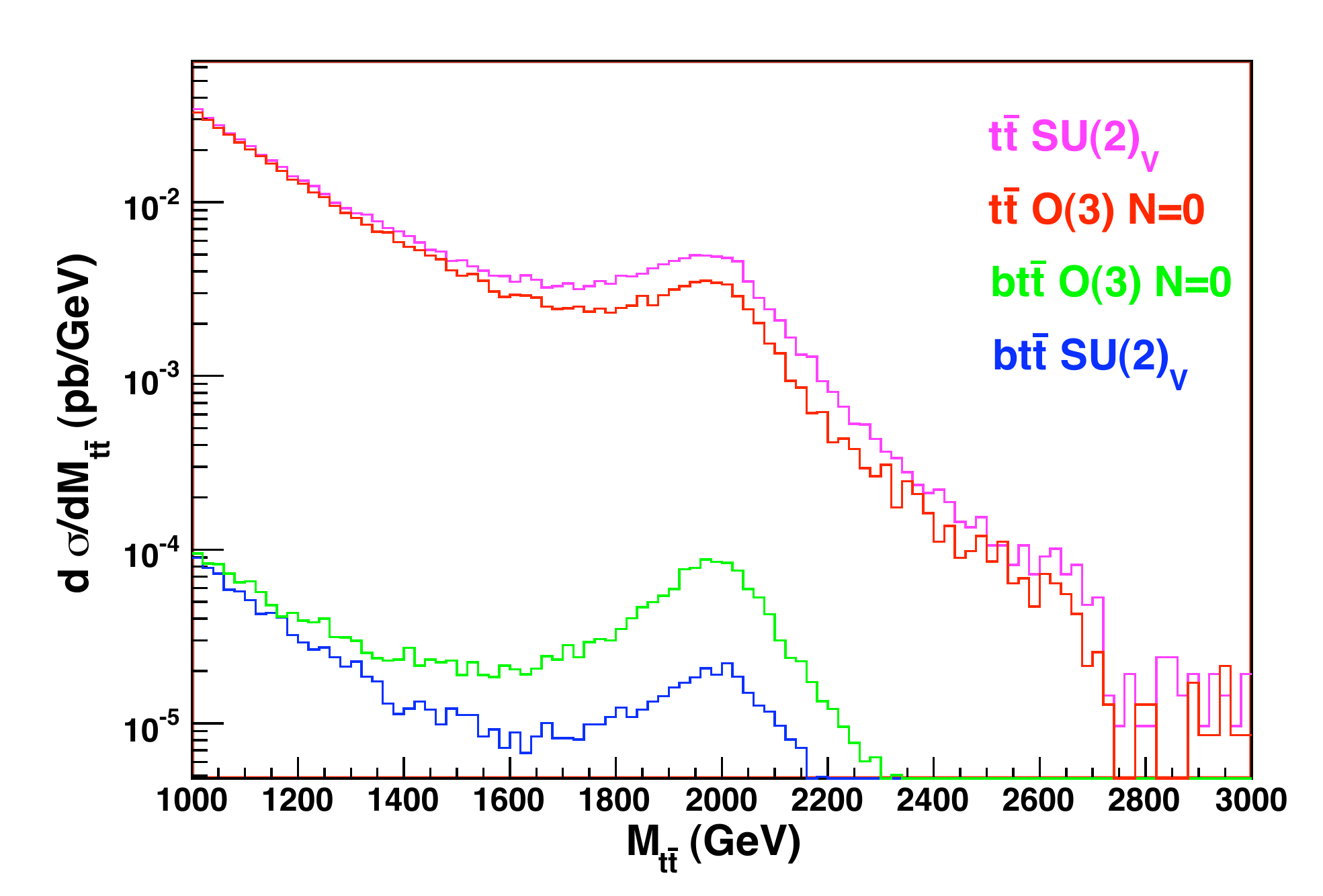} 
\caption{\label{fig:diffcrosplot}
Invariant mass distribution of $t\bar t$ in the standard RS model 
($\rm{SU(2)_V}$ custodial symmetry) and the model with a larger 
$\rm{O(3)}$ custodial symmetry in $pp \to t\bar t$ and $pp \to b t\bar t$ 
respectively.
}
\end{figure}

As indicated above, models with the extra custodial symmetry to protect 
the  $Z$-$b_L$-$b_L$ coupling from large corrections have considerably more
freedom to locate $Q_3$ closer to the IR brane, and considerations of the
$T$ parameter prefer to do so.  This enhances the $g^1$ coupling to
left-handed bottoms (up to about $3 g_S$) and results in large
production from bottom quark fusion, as shown in Figure~\ref{fig:xsec}.
It would be useful to be able to discern that the increase over the
expected production rate in the standard bulk SM RS picture is because
of this enhancement of the coupling to bottom (which would be suggestive
of the expanded custodial symmetry), as opposed to
a straight enhancement of the coupling to all light quarks (which would
be more suggestive of a large kinetic term on the IR boundary).
One could study the rapidity distribution of the $g^1$ itself (as reflected
in the final state top pair distribution).  The fact that both
$b$ and $\bar{b}$ are sea quarks would imply a more central rapidity 
distribution than would result from $q$ and $\bar{q}$, because $q$ as a
valence quark will tend to carry more momentum than $\bar{q}$. 
However, the $g^1$ rapidity distribution is only modestly sensitive
to the initial state, and is also sensitive to the $g^1$ mass and width.
Thus, we turn to a more straight-forward measure of the contribution of
$b \bar{b}$ to $g^1$ production \footnote{We are grateful to Tao Han for this
suggestion.} which is to compare the rate of $g^1 \rightarrow t \bar{t}$ 
to $b g^1 \rightarrow b t \bar{t}$.  In Figure~\ref{fig:bxsec}, we
present these rates for standard RS, the model with $\kappa r_{IR}=5$, and
the model with $Q_3$ localized around the IR brane.  We find that as expected,
the rate for the model with custodial symmetry is enhanced by the large bottom
coupling by about an order of magnitude.  
In addition, the model with IR boundary kinetic terms shows a rate which is 
suppressed by a factor of about five, because while the boundary
kinetic term slightly enhances the coupling of the UV-localized $b_R$, it more
dramatically suppresses the coupling to the IR-localized $b_L$
(c.f. Figure~\ref{fig:gffbkcoup}). 
Ultimately, one must include the SM background and detector efficiencies for 
a specific decay channel of $g^1$. As a step in this direction,
in Figure~\ref{fig:diffcrosplot} we plot the differential cross-section 
for both the $pp \to t\bar t$ and $pp \to b t\bar t$ signals
and SM backgrounds with respect to the $t \bar{t}$ invariant mass, 
in the standard RS model and one with a larger custodial symmetry. 
In both cases, for $M_{g^1} = 2$~TeV, a peak is visible above the SM
background, and the size of $g^1 b$ production relative to $g^1$ production
discriminates between the two models.

The width of $g^1$ is strongly dominated by the states close to the IR brane
to which it couples strongly.  Generically, the partial width into 
$f \bar{f}$ for which the left- and right-chiral interactions with $g^1$
are $g_L$ and $g_R$ is given by,
\begin{eqnarray}
\Gamma_{G1 \rightarrow f \bar{f}} & = & \frac{1}{48 \pi M_{g^1}}
\sqrt{1 - \frac{4 m^2_f}{M^2_{g^1}}} 
\left[ \left( g_L^2 + g_R^2 \right) \left(M_{g^1}^2-m_f^2 \right)
+ 6 g_{L} g_{R} m_f^2 \right] \nonumber \\
 & \simeq & \frac{M_{g^1}}{48 \pi} \left( g_L^2 + g_R^2 \right) ~,
\label{decayequation}
\end{eqnarray}
where the final approximation holds in the limit $M_{g^1} \gg m_f$.
Decays to top quarks are always important, because either $t_R$ or $Q_3$
must be IR-localized to realize the large top mass.  In addition, when the
custodial partner KK quarks are light enough for $g^1$ to decay into them,
they will also take a substantial fraction of the branching ratio, because
they are also IR-localized and have large coupling.  The IR boundary
kinetic terms can suppress the coupling to top, and enhance the decay into
light quarks.  In Table~\ref{tab:brs} we list the branching ratios into
top quarks, bottom quarks, light quarks (jets) and exotic quarks in several
different RS models. 
The total width also sensitively depends on the couplings, and how many
custodial partners are available as decay modes.  The width is generally
large, owing to the strong couplings present, and it may be possible to
reconstruct it from the final state invariant mass distributions, which
would also allow one to use it as an additional source of information.
The final column of Table~\ref{tab:brs} shows the total
width $\Gamma_{g^1} / M_{g^1}$ for each model.  Variations are typically
around $5\%$, with the exception of the model with an extra custodial
partner, whose very strong coupling has a big effect on the width.
In fact, allowing too many additional custodial partners will rapidly 
drive $\Gamma_{g^1} \gtrsim M_{g^1}$, an indication of a break-down of
perturbation theory.  From Eq.~(\ref{decayequation}), we can infer that there
can be at most four new custodial quarks whose masses are less than
$M_{g^1} / 2$.

\begin{table}
\begin{tabular}{l|ccccc}
Model           & ~~top quarks~~ & ~~bottom quarks~~ & 
~~light quarks~~ & ~~custodial partners~~ & ~~$\Gamma_{g^1}/M_{g^1}$~~ \\
\hline
Basic RS~~~~~~          & 92.6\% & 5.7\%  & 1.7\%  & & 0.14 \\
$\kappa r_{IR} = 5$     & 2.6\%  & 13.2\% & 84.2\% & & 0.11 \\
$\kappa r_{IR} = 20$    & 7.8\%  & 15.1\% & 77.1\% & & 0.05 \\
$O(3)$, $N=0$           & 48.8\% & 49.0\% & 2.0\%  & & 0.11 \\
$O(3)$, $N=1$           & 14.6\% & 14.6\% & 0.6\%  & 70.2\% & 0.40 \\
\end{tabular}
\caption{\label{tab:brs} The branching ratios of $g^1$ into
tops, bottoms, light quarks (jets), and custodial partners, 
as well as the total width $\Gamma_{g^1} / M_{g^1}$, 
for several different RS scenarios in the limit $M_{g^1} \gg m_f$.}
\end{table}

\begin{figure}[tbp]
\hspace*{-0.75cm}
\includegraphics[angle=0,scale=0.65]{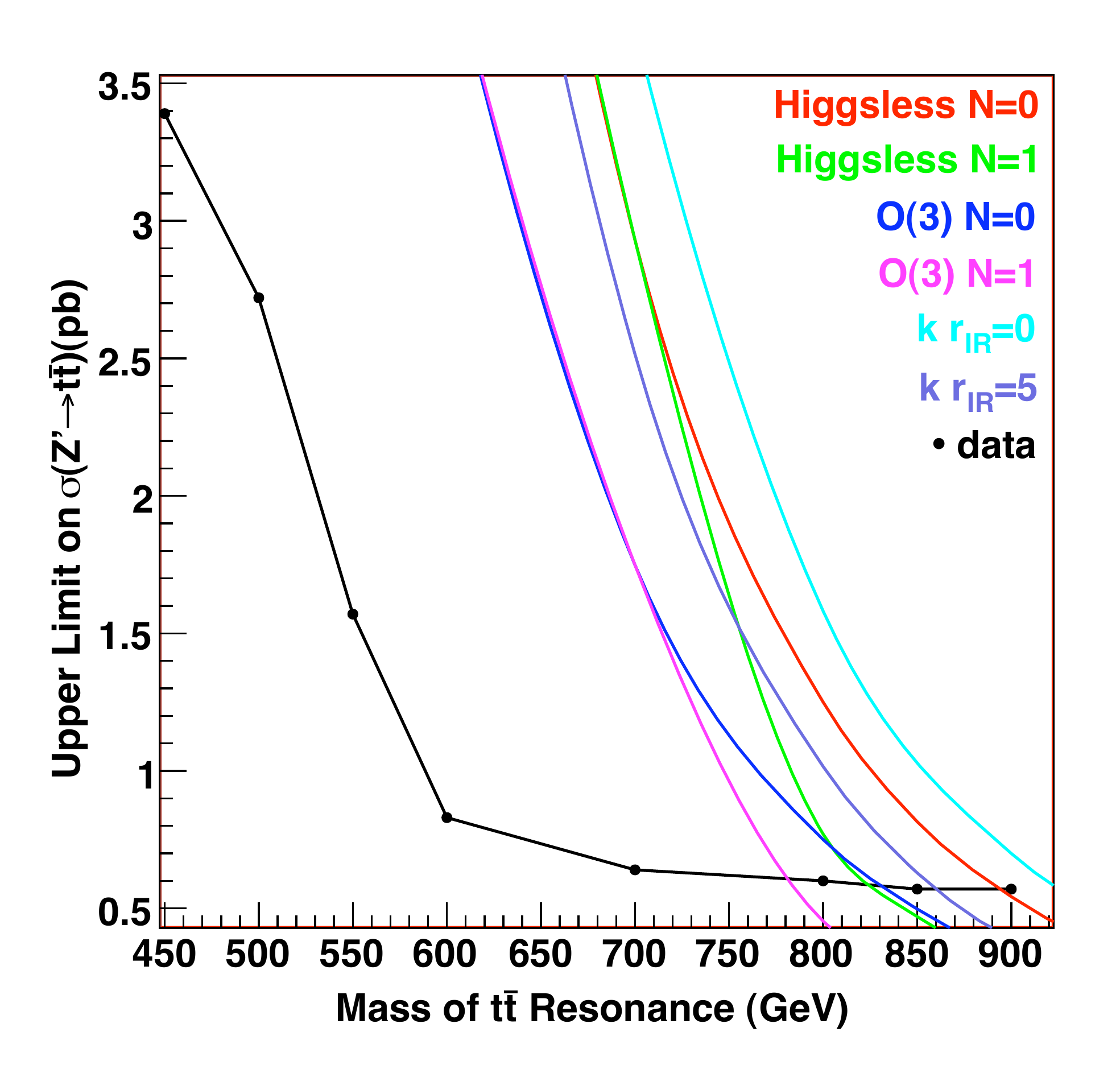} 
\caption{\label{fig:tevatron}
Cross section for $p \bar{p} \rightarrow g^1 \rightarrow t \bar{t}$ at the
Tevatron as a function of the mass of $g^1$, 
compared with the CDF exclusion curve. The mass of custodial partners is 360GeV.}
\end{figure} 

\begin{figure}[tpb]
\hspace*{-1.8cm}
\includegraphics[angle=0,scale=0.85]{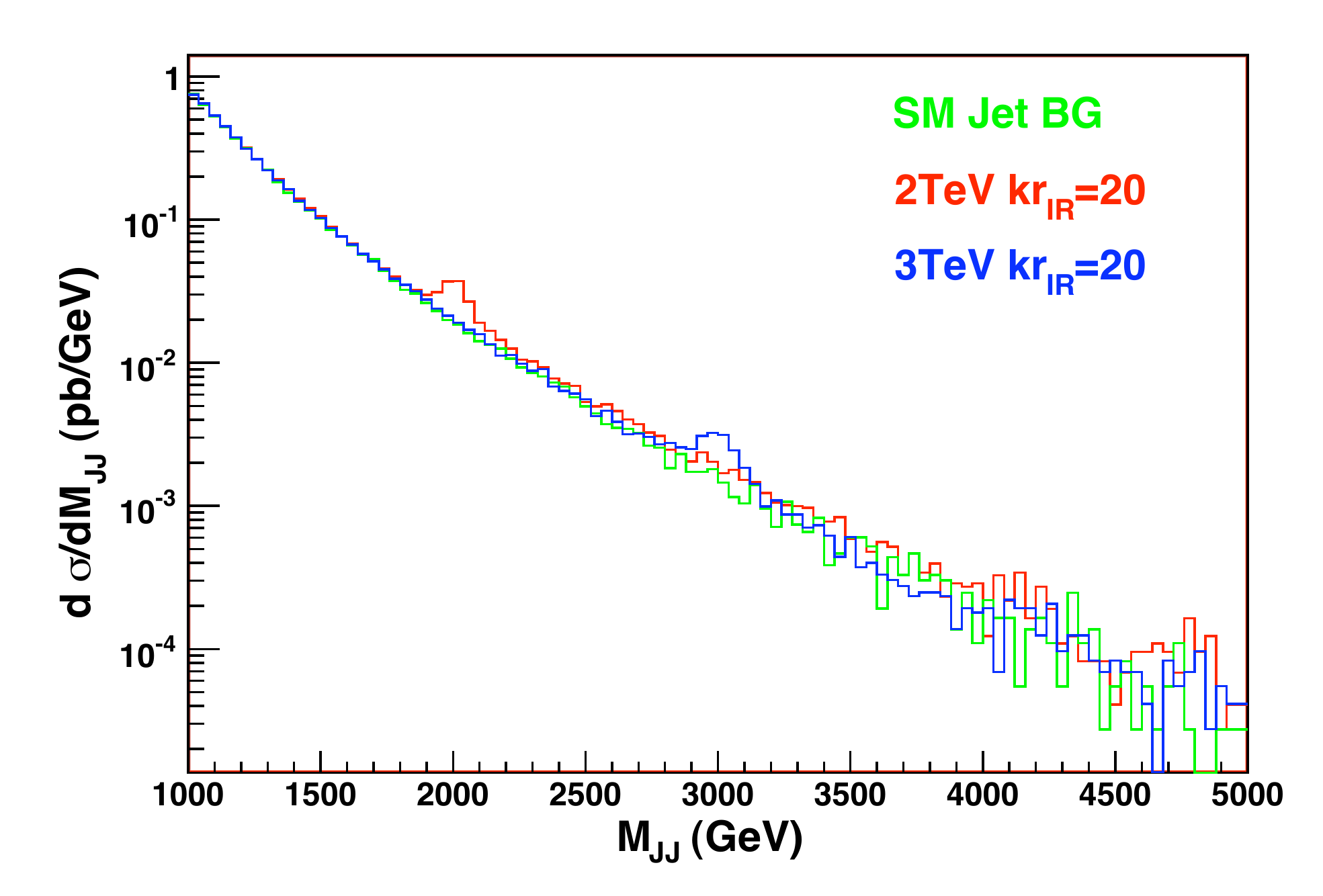} 
\caption{\label{fig:jetplot}
Invariant mass distribution of QCD dijets coming from the KK gluon 
resonance in models with large brane kinetic term ($\kappa r_{IR} = 20$), 
and the SM prediction.  The cuts $p_T >20$GeV, $|\eta| <1.0$, and 
invariant mass $>1$TeV are applied. 
}
\end{figure}

In models with large boundary kinetic terms, $g^1$ primarily decays 
into light quarks, swamping the decay into tops, and its over-all width
becomes much narrower.  This fact, combined with the enhancement of
$g^1$ production, allows for the possibility that one could discover
$g^1$ in the dijet mode, against the large QCD background.  To explore
this possibility, in Figure~\ref{fig:jetplot} we plot the invariant mass 
distribution of QCD dijets (with rough acceptance cuts 
$|\eta| <1.0$ and $p_T > 20$~GeV to reduce the SM background). 
For $M_{g^1} = 2$ or $3$ TeV, we can reconstruct a peak against the dijet
background with ample statistics.  Based on the size of the signal and
background, we estimate that one could potentially discover $g^1$ even if
its mass is larger than 4~TeV in such models.

The highly chiral nature of the couplings of $g^1$ to top, bottom,
or the custodial partners may be visible as an observable
\cite{Agashe:2006hk}.  The top final
state is particularly promising, because the left-handed nature of the 
$W$-$t$-$b$ interaction implies that the top decay automatically analyzes its
production polarization.  For example, the standard RS scenario has
about $95\%$ decays into right-polarized tops, whereas the model with
$\kappa r_{IR} = 10$ has roughly equal decays into left- and right-polarized
tops, and the model with expanded custodial symmetry with
$Q_3$ localized at the IR brane has about $99\%$ decays into left-polarized
tops.

Finally, given the large cross-sections, it is natural to ask what the 
current bounds from the Tevataron on anomalous top production imply for the 
KK gluon mass. A recent analysis from CDF \cite{tevbound} has set bounds on 
narrow resonances in the $t\bar t$ invariant mass spectrum. While the 
analysis does not strictly apply in this case, since the KK gluon is wider 
than the machine resolution, the actual bound will be close to that quoted 
in the analysis. We have plotted this in Fig. \ref{fig:tevatron}, along 
with representative cross-sections from the models under investigation here. 
Note that this excludes Higgsless models with KK masses below about 
850~GeV, and that includes the region favored by unitarity 
in $WW$ scattering.

\section{Interference}
\label{sec:interfere}

\begin{figure}[b]
\includegraphics[angle=0,scale=1]{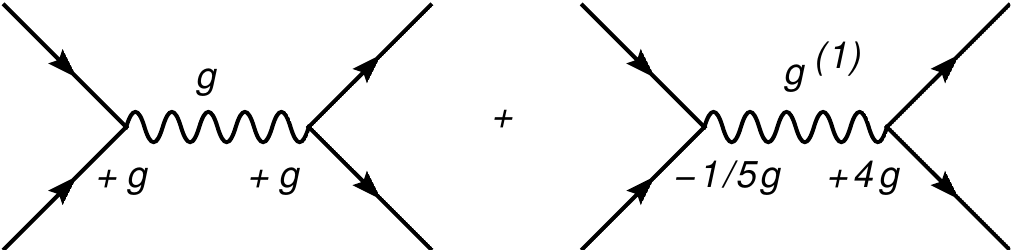} 
\caption{\label{fig:feyn}
Graphs that interfere allowing measurement of the sign of the light quark 
coupling.
}
\end{figure}

There is an intriguing feature of the fermion couplings to $g^1$: 
the sign of the coupling depends on the sign of the $g^1$ wave function close
to where the fermion is localized.  As a KK mode, the $g^1$ wave function
contains a node, and changes sign from one side of the extra dimension to the
other.  As a result the UV fermions 
have a minus sign relative to the zero mode gluon coupling, 
while the IR fermions have a plus sign. This sign should be visible in the 
interference between $s$-channel gluon and KK-gluon production of $t\bar t$, 
as illustrated in Fig. \ref{fig:feyn}.

\begin{figure}
\hspace*{-1.8cm}
\includegraphics[angle=0,scale=0.9]{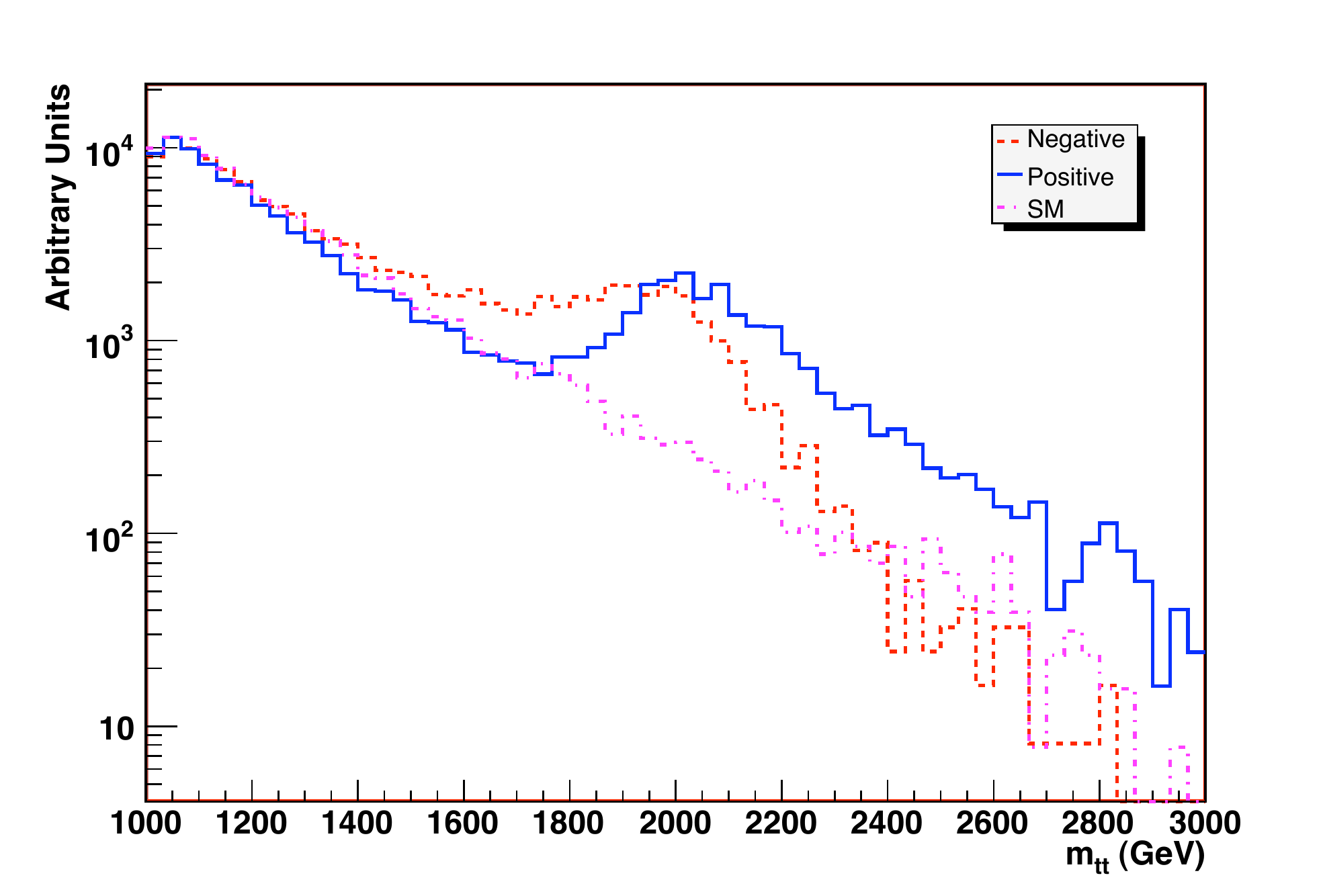} 
\caption{\label{fig:intplot}
Invariant mass distribution of $pp\to t\bar t$ in models with positive and 
negative coupling to light fermions, along with the SM prediction.
}
\end{figure}

To quantify this effect we propose an asymmetry parameter $A_{i}$. This 
parameter should be positive or negative depending on the sign of the light 
quark coupling and be zero in the Standard Model. We accomplish this with the 
definition
\begin{gather}
A_{i} = -
\frac{\int dm (\frac{d\sigma}{dm} - 
\frac{d\sigma}{dm}_{\rm SM})*\Theta(m - M_{g^1})}
{\int dm |\frac{d\sigma}{dm} - \frac{d\sigma}{dm}_{SM}|}.
\end{gather}
Here $m$ is the invariant mass in the $t\bar t$ distribution and $M_{g^1}$ is 
the center of the resonance. The logic of this choice is that: {\it i.} The 
SM contribution is subtracted to determine if the interference is positive or 
negative; {\it ii.} the sign of the interference changes as the resonance is 
crossed, hence the $\Theta$-function; {\it iii.} As is well-known, a 
positive sign will produce negative interference below the resonance and 
positive above due to the sign of the resonance propagator 
$1 / (s - M_{g^1}^2)$, hence the overall minus sign. With this definition the 
sign of $A_{i}$ will be that of the light quark coupling.

The normalization of the data with respect to the SM calculation is 
problematic. Since the resonance will result in a much larger overall 
cross-section, one should not normalize to the total number of events. We 
choose to normalize to the lowest-mass bin used in calculating the 
asymmetry, which allows extraction of the normalization from data, while 
retaining all available information in the region near the resonance.

We present values of $A_{i}$ for several masses in the basic RS model in 
Table~\ref{tab:asym}. We also show the value obtained by switching the sign 
of the light quark coupling. We have included a crude estimate of the 
smearing by shifting the value of the top and anti-top 4-momentum by a 
gaussian random number with width given by the ATLAS jet resolution. 
Since the uncertainty in top reconstruction will be dominated by the 
jet uncertainty this gives the correct order-of-magnitude for the smearing;
we leave more refined estimates for future work.
We find that the smearing 
makes little difference, as the resonance width is larger 
than the detector resolution. The results in Table~\ref{tab:asym} indicate 
that if a resonance is observed in $t\bar t$ production, $A_{i}$ is a 
promising variable to extract information about the underlying theory.

\begin{table}
\begin{tabular}{l|cc}
$g^{(1)}$ Mass           & ~~plus~~ & ~~minus~~ \\
\hline
2 TeV          & 0.57 & -0.44 \\
3 TeV          & 0.54 & -0.28 \\
4 TeV          & 0.52 & -0.16 \\
\end{tabular}
\caption{\label{tab:asym} 
Asymmetry parameter $A_{i}$ for $t\bar t$ resonances with negative 
(corresponding to basic RS) and positive light quark couplings.
}
\end{table}

\section{Conclusions}
\label{sec:conclusions}

We have investigated the structure of the KK gluon resonance in several 
variants of the RS model. We find that this structure contains information 
that will help to distinguish between models even in the absence of data 
from the electroweak sector.
The width and branching ratios will constrain the location of the fermion 
zero-modes as well as the number of light KK modes into which the KK gluon 
can decay. In addition, the ratio of cross-sections for producing 
the $g^{1}$ directly and in association with a $b$-jet will give specific 
information about the localization of the third generation quarks. 
Specifically, a large coupling to $b \bar b$ will prefer a model 
where the $Z\to b\bar b$ vertex is protected by an extended custodial symmetry.
In some models, with large boundary kinetic terms, the $g^1$ can primarily
decay into dijets, and it seems promising that in such models one can
discern $g^1$ against the large QCD background up to masses somewhat larger
than 4 TeV.

Finally, we find that the relative sign of the coupling to light quarks and 
to tops can be measured in the interference with $s$-channel gluon exchange. 
This provides an important consistancy check on the overall picture of the 
fermion geography and the mechanism by which flavor hierarchies are realized
in the fermion Yukawa couplings.

The discovery of $g^1$ is an important first step in the discovery of RS, and
further observables such as its production rate, associated rate with bottom
quarks, total width and branching ratios, and interference with SM $t \bar{t}$
production, can yield information about the nature of the the RS construction,
and the parameters which describe it.

{\bf Acknowledgements }

The authors would like to thank Erik Brubaker, Giacomo Cacciapaglia, Tao Han, Gregory House, 
Tom LeCompte, David Mckeen, Arjun Menon, Jose Santiago,
and Carlos Wagner for helpful discussions.
Research at Argonne National Laboratory is 
supported in part by the Department of Energy 
under contract DE-AC02-06CH11357.


\newpage

\begin{thebibliography}{99}

%\cite{Randall:1999vf}
\bibitem{Randall:1999vf}
  L.~Randall and R.~Sundrum,
  %``An alternative to compactification,''
  Phys.\ Rev.\ Lett.\  {\bf 83}, 4690 (1999)
  [arXiv:hep-th/9906064].
  L.~Randall and R.~Sundrum,
  %``A large mass hierarchy from a small extra dimension,''
  Phys.\ Rev.\ Lett.\  {\bf 83}, 3370 (1999)
  [arXiv:hep-ph/9905221].
  %%CITATION = HEP-PH 9905221;%%

\bibitem{jmr}
These arguments were lucidly put together my J. March-Russell, talk at 
HCP 2007, Isola d'Elba, Italy.

%\cite{Arkani-Hamed:2000ds}
\bibitem{Arkani-Hamed:2000ds}
  N.~Arkani-Hamed, M.~Porrati and L.~Randall,
  %``Holography and phenomenology,''
  JHEP {\bf 0108}, 017 (2001)
  [arXiv:hep-th/0012148].
  %%CITATION = JHEPA,0108,017;%%

%\cite{Randall:2001gb}
\bibitem{Randall:2001gb}
  L.~Randall and M.~D.~Schwartz,
  %``Quantum field theory and unification in AdS5,''
  JHEP {\bf 0111}, 003 (2001)
  [arXiv:hep-th/0108114];
  %%CITATION = JHEPA,0111,003;%%
%\cite{Agashe:2002pr}
%\bibitem{Agashe:2002pr}
  K.~Agashe, A.~Delgado and R.~Sundrum,
  %``Grand unification in RS1,''
  Annals Phys.\  {\bf 304}, 145 (2003)
  [arXiv:hep-ph/0212028];
  %%CITATION = APNYA,304,145;%%
%\cite{Carena:2003fx}
%\bibitem{Carena:2003fx}
  M.~Carena, A.~Delgado, E.~Ponton, T.~M.~P.~Tait and C.~E.~M.~Wagner,
  %``Precision electroweak data and unification of couplings in warped extra
  %dimensions,''
  Phys.\ Rev.\  D {\bf 68}, 035010 (2003)
  [arXiv:hep-ph/0305188];
  %%CITATION = PHRVA,D68,035010;%%
%\cite{Agashe:2005vg}
%\bibitem{Agashe:2005vg}
  K.~Agashe, R.~Contino and R.~Sundrum,
  %``Top compositeness and precision unification,''
  Phys.\ Rev.\ Lett.\  {\bf 95}, 171804 (2005)
  [arXiv:hep-ph/0502222].
  %%CITATION = PRLTA,95,171804;%%

%\cite{Huber:2000ie}
\bibitem{Huber:2000ie}
  S.~J.~Huber and Q.~Shafi,
  %``Fermion masses, mixings and proton decay in a Randall-Sundrum model,''
  Phys.\ Lett.\  B {\bf 498}, 256 (2001)
  [arXiv:hep-ph/0010195].
  %%CITATION = PHLTA,B498,256;%%

%\cite{Agashe:2004ci}
\bibitem{Agashe:2004ci}
  K.~Agashe and G.~Servant,
  %``Warped unification, proton stability and dark matter,''
  Phys.\ Rev.\ Lett.\  {\bf 93}, 231805 (2004)
  [arXiv:hep-ph/0403143];
  %%CITATION = PRLTA,93,231805;%%
%\cite{Agashe:2004bm}
%\bibitem{Agashe:2004bm}
  K.~Agashe and G.~Servant,
  %``Baryon number in warped GUTs: Model building and (dark matter related)
  %phenomenology,''
  JCAP {\bf 0502}, 002 (2005)
  [arXiv:hep-ph/0411254].
  %%CITATION = JCAPA,0502,002;%%

%\cite{Davoudiasl:1999tf}
\bibitem{Davoudiasl:1999tf}
  H.~Davoudiasl, J.~L.~Hewett and T.~G.~Rizzo,
  %``Bulk gauge fields in the Randall-Sundrum model,''
  Phys.\ Lett.\  B {\bf 473}, 43 (2000)
  [arXiv:hep-ph/9911262];
  %%CITATION = PHLTA,B473,43;%%
%\cite{Pomarol:1999ad}
%\bibitem{Pomarol:1999ad}
A.~Pomarol,
%``Gauge bosons in a five-dimensional theory with localized gravity,''
Phys.\ Lett.\  B {\bf 486}, 153 (2000)
[arXiv:hep-ph/9911294].
%%CITATION = PHLTA,B486,153;%%
%\cite{Csaki:2002gy}
%\bibitem{Csaki:2002gy}
  C.~Csaki, J.~Erlich and J.~Terning,
  %``The effective Lagrangian in the Randall-Sundrum model and electroweak
  %physics,''
  Phys.\ Rev.\  D {\bf 66}, 064021 (2002)
  [arXiv:hep-ph/0203034].
  %%CITATION = PHRVA,D66,064021;%%

%\cite{Agashe:2003zs}
\bibitem{Agashe:2003zs}
  K.~Agashe, A.~Delgado, M.~J.~May and R.~Sundrum,
  %``RS1, custodial isospin and precision tests,''
  JHEP {\bf 0308}, 050 (2003)
  [arXiv:hep-ph/0308036].
  %%CITATION = JHEPA,0308,050;%%

%\cite{Davoudiasl:2002ua}
\bibitem{Davoudiasl:2002ua}
  H.~Davoudiasl, J.~L.~Hewett and T.~G.~Rizzo,
  %``Brane localized kinetic terms in the Randall-Sundrum model,''
  Phys.\ Rev.\  D {\bf 68}, 045002 (2003)
  [arXiv:hep-ph/0212279];
  %%CITATION = PHRVA,D68,045002;%%
%\cite{Carena:2002dz}
%\bibitem{Carena:2002dz}
  M.~Carena, E.~Ponton, T.~M.~P.~Tait and C.~E.~M.~Wagner,
  %``Opaque branes in warped backgrounds,''
  Phys.\ Rev.\  D {\bf 67}, 096006 (2003)
  [arXiv:hep-ph/0212307];
  %%CITATION = PHRVA,D67,096006;%%
%\cite{Carena:2004zn}
%\bibitem{Carena:2004zn}
  M.~Carena, A.~Delgado, E.~Ponton, T.~M.~P.~Tait and C.~E.~M.~Wagner,
  %``Warped fermions and precision tests,''
  Phys.\ Rev.\  D {\bf 71}, 015010 (2005)
  [arXiv:hep-ph/0410344].
  %%CITATION = PHRVA,D71,015010;%%

%\cite{Agashe:2006at}
\bibitem{Agashe:2006at}
  K.~Agashe, R.~Contino, L.~Da Rold and A.~Pomarol,
  %``A custodial symmetry for Z b anti-b,''
  Phys.\ Lett.\ B {\bf 641}, 62 (2006)
  [arXiv:hep-ph/0605341].
  %%CITATION = HEP-PH 0605341;%%

%\cite{Agashe:2006hk}
\bibitem{Agashe:2006hk}
  K.~Agashe, A.~Belyaev, T.~Krupovnickas, G.~Perez and J.~Virzi,
  %``LHC signals from warped extra dimensions,''
  arXiv:hep-ph/0612015.
  %%CITATION = HEP-PH 0612015;%%
  B.~Lillie, L.~Randall and L.~T.~Wang,
  %``The Bulk RS KK-gluon at the LHC,''
  arXiv:hep-ph/0701166.
  %%CITATION = HEP-PH 0701166;%%

%\cite{Contino:2006qr}
\bibitem{Contino:2006qr}
  R.~Contino, L.~Da Rold and A.~Pomarol,
  %``Light custodians in natural composite Higgs models,''
  arXiv:hep-ph/0612048.
  %%CITATION = HEP-PH 0612048;%%

%\cite{Carena:2007ua}
\bibitem{Carena:2007ua}
  M.~Carena, E.~Ponton, J.~Santiago and C.~E.~M.~Wagner,
  %``Electroweak constraints on warped models with custodial symmetry,''
  arXiv:hep-ph/0701055.
  %%CITATION = HEP-PH 0701055;%%

%\cite{Grossman:1999ra}
\bibitem{Grossman:1999ra}
Y.~Grossman and M.~Neubert,
%``Neutrino masses and mixings in non-factorizable geometry,''
Phys.\ Lett.\  B {\bf 474}, 361 (2000)
  [arXiv:hep-ph/9912408].
%%CITATION = PHLTA,B474,361;%%

\bibitem{Carena:2002me}
  M.~Carena, T.~M.~P.~Tait and C.~E.~M.~Wagner,
  %``Branes and orbifolds are opaque,''
  Acta Phys.\ Polon.\  B {\bf 33}, 2355 (2002)
  [arXiv:hep-ph/0207056].
  %%CITATION = APPOA,B33,2355;%%

%\cite{Georgi:2000ks}
\bibitem{Georgi:2000ks}
  H.~Georgi, A.~K.~Grant and G.~Hailu,
  %``Brane couplings from bulk loops,''
  Phys.\ Lett.\  B {\bf 506}, 207 (2001)
  [arXiv:hep-ph/0012379].
  %%CITATION = PHLTA,B506,207;%%

%\cite{Choudhury:2001hs}
\bibitem{Choudhury:2001hs}
  D.~Choudhury, T.~M.~P.~Tait and C.~E.~M.~Wagner,
  %``Beautiful mirrors and precision electroweak data,''
  Phys.\ Rev.\  D {\bf 65}, 053002 (2002)
  [arXiv:hep-ph/0109097].
  %%CITATION = PHRVA,D65,053002;%%

%\cite{Agashe:2004rs}
\bibitem{Agashe:2004rs}
  K.~Agashe, R.~Contino and A.~Pomarol,
  %``The minimal composite Higgs model,''
  Nucl.\ Phys.\  B {\bf 719}, 165 (2005)
  [arXiv:hep-ph/0412089].
  %%CITATION = NUPHA,B719,165;%%

%\cite{Carena:2006bn}
\bibitem{Carena:2006bn}
  M.~Carena, E.~Ponton, J.~Santiago and C.~E.~M.~Wagner,
  %``Light Kaluza-Klein states in Randall-Sundrum models with custodial
  %SU(2),''
  Nucl.\ Phys.\  B {\bf 759}, 202 (2006)
  [arXiv:hep-ph/0607106].
  %%CITATION = NUPHA,B759,202;%%
%\cite{Medina:2007hz}
%\bibitem{Medina:2007hz}
A.~D.~Medina, N.~R.~Shah and C.~E.~M.~Wagner,
%``Gauge-Higgs Unification and Radiative Electroweak Symmetry Breaking in
%Warped Extra Dimensions,''
arXiv:0706.1281 [hep-ph].
%%CITATION = ARXIV:0706.1281;%%

%\cite{Djouadi:2006rk}
\bibitem{Djouadi:2006rk}
  A.~Djouadi, G.~Moreau and F.~Richard,
  %``Resolving the A(FB)(b) puzzle in an extra dimensional model with an
  %extended gauge structure,''
  Nucl.\ Phys.\  B {\bf 773}, 43 (2007)
  [arXiv:hep-ph/0610173].
  %%CITATION = NUPHA,B773,43;%%

\bibitem{lepewwg}
See {\tt http://lepewwg.web.cern.ch/LEPEWWG/} for recent fits to $S$ and $T$.

%\cite{Csaki:2003dt}
\bibitem{Csaki:2003dt}
  C.~Csaki, C.~Grojean, H.~Murayama, L.~Pilo and J.~Terning,
  %``Gauge theories on an interval: Unitarity without a Higgs,''
  Phys.\ Rev.\  D {\bf 69}, 055006 (2004)
  [arXiv:hep-ph/0305237].
  %%CITATION = PHRVA,D69,055006;%%
%\cite{Csaki:2003zu}
%\bibitem{Csaki:2003zu}
  C.~Csaki, C.~Grojean, L.~Pilo and J.~Terning,
  %``Towards a realistic model of Higgsless electroweak symmetry breaking,''
  Phys.\ Rev.\ Lett.\  {\bf 92}, 101802 (2004)
  [arXiv:hep-ph/0308038].
  %%CITATION = PRLTA,92,101802;%%
%\cite{Cacciapaglia:2004jz}
%\bibitem{Cacciapaglia:2004jz}
  G.~Cacciapaglia, C.~Csaki, C.~Grojean and J.~Terning,
  %``Oblique corrections from Higgsless models in warped space,''
  Phys.\ Rev.\  D {\bf 70}, 075014 (2004)
  [arXiv:hep-ph/0401160].
  %%CITATION = PHRVA,D70,075014;%%
%\cite{Cacciapaglia:2004rb}
%\bibitem{Cacciapaglia:2004rb}
  G.~Cacciapaglia, C.~Csaki, C.~Grojean and J.~Terning,
  %``Curing the ills of Higgsless models: The S parameter and unitarity,''
  Phys.\ Rev.\  D {\bf 71}, 035015 (2005)
  [arXiv:hep-ph/0409126].
  %%CITATION = PHRVA,D71,035015;%%
%\cite{Cacciapaglia:2005pa}
%\bibitem{Cacciapaglia:2005pa}
  G.~Cacciapaglia, C.~Csaki, C.~Grojean, M.~Reece and J.~Terning,
  %``Top and bottom: A brane of their own,''
  Phys.\ Rev.\  D {\bf 72}, 095018 (2005)
  [arXiv:hep-ph/0505001].
  %%CITATION = PHRVA,D72,095018;%%
%\cite{Csaki:2003sh}
%\bibitem{Csaki:2003sh}
  C.~Csaki, C.~Grojean, J.~Hubisz, Y.~Shirman and J.~Terning,
  %``Fermions on an interval: Quark and lepton masses without a Higgs,''
  Phys.\ Rev.\  D {\bf 70}, 015012 (2004)
  [arXiv:hep-ph/0310355].
  %%CITATION = PHRVA,D70,015012;%%
%\cite{Nomura:2003du}
%\bibitem{Nomura:2003du}
  Y.~Nomura,
  %``Higgsless theory of electroweak symmetry breaking from warped space,''
  JHEP {\bf 0311}, 050 (2003)
  [arXiv:hep-ph/0309189].
  %%CITATION = JHEPA,0311,050;%%
%\cite{Barbieri:2003pr}
%\bibitem{Barbieri:2003pr}
  R.~Barbieri, A.~Pomarol and R.~Rattazzi,
  %``Weakly coupled Higgsless theories and precision electroweak tests,''
  Phys.\ Lett.\  B {\bf 591}, 141 (2004)
  [arXiv:hep-ph/0310285].
  %%CITATION = PHLTA,B591,141;%%
%\cite{Davoudiasl:2003me}
%\bibitem{Davoudiasl:2003me}
  H.~Davoudiasl, J.~L.~Hewett, B.~Lillie and T.~G.~Rizzo,
   ``Higgsless electroweak symmetry breaking in warped backgrounds:  Constraints
  %and signatures,''
  Phys.\ Rev.\  D {\bf 70}, 015006 (2004)
  [arXiv:hep-ph/0312193].
  %%CITATION = PHRVA,D70,015006;%%
%\cite{Davoudiasl:2004pw}
%\bibitem{Davoudiasl:2004pw}
  H.~Davoudiasl, J.~L.~Hewett, B.~Lillie and T.~G.~Rizzo,
  %``Warped Higgsless models with IR-brane kinetic terms,''
  JHEP {\bf 0405}, 015 (2004)
  [arXiv:hep-ph/0403300].
  %%CITATION = JHEPA,0405,015;%%
%\cite{Hewett:2004dv}
%\bibitem{Hewett:2004dv}
  J.~L.~Hewett, B.~Lillie and T.~G.~Rizzo,
  %``Monte Carlo exploration of warped Higgsless models,''
  JHEP {\bf 0410}, 014 (2004)
  [arXiv:hep-ph/0407059].
  %%CITATION = JHEPA,0410,014;%%
%\cite{Chivukula:2004pk}
%\bibitem{Chivukula:2004pk}
  R.~S.~Chivukula, E.~H.~Simmons, H.~J.~He, M.~Kurachi and M.~Tanabashi,
   ``The structure of corrections to electroweak interactions in Higgsless
  %models,''
  Phys.\ Rev.\  D {\bf 70}, 075008 (2004)
  [arXiv:hep-ph/0406077].
  %%CITATION = PHRVA,D70,075008;%%
%\cite{Chivukula:2004af}
%\bibitem{Chivukula:2004af}
  R.~S.~Chivukula, E.~H.~Simmons, H.~J.~He, M.~Kurachi and M.~Tanabashi,
  %``Universal non-oblique corrections in Higgsless models and beyond,''
  Phys.\ Lett.\  B {\bf 603}, 210 (2004)
  [arXiv:hep-ph/0408262].
  %%CITATION = PHLTA,B603,210;%%

%\cite{Cacciapaglia:2006mz}
\bibitem{Cacciapaglia:2006mz}
  G.~Cacciapaglia, C.~Csaki, G.~Marandella and J.~Terning,
  %``The gaugephobic Higgs,''
  JHEP {\bf 0702}, 036 (2007)
  [arXiv:hep-ph/0611358].
  %%CITATION = JHEPA,0702,036;%%
%\cite{Cacciapaglia:2006gp}
%\bibitem{Cacciapaglia:2006gp}
  G.~Cacciapaglia, C.~Csaki, G.~Marandella and J.~Terning,
  %``A new custodian for a realistic Higgsless model,''
  Phys.\ Rev.\  D {\bf 75}, 015003 (2007)
  [arXiv:hep-ph/0607146].
  %%CITATION = PHRVA,D75,015003;%%

%\cite{Sjostrand:2006za}
\bibitem{Sjostrand:2006za}
  T.~Sjostrand, S.~Mrenna and P.~Skands,
  %``PYTHIA 6.4 physics and manual,''
  JHEP {\bf 0605}, 026 (2006)
  [arXiv:hep-ph/0603175].
  %%CITATION = JHEPA,0605,026;%%

%\cite{Alwall:2007st}
\bibitem{Alwall:2007st}
J.~Alwall {\it et al.},
%``MadGraph/MadEvent v4: The New Web Generation,''
  arXiv:0706.2334 [hep-ph].
%%CITATION = ARXIV:0706.2334;%%

\bibitem{tevbound} M. Kagan, D. Amidei, C. Cully, T. Schwarz, and M. Soderberg 
http://www-cdf.fnal.gov/physics/new/top/2006/mass/mttb/pub\_page.html


\end{thebibliography}
\end{document}